# Photonics-assisted analog wideband self-interference cancellation for in-band full-duplex MIMO systems with adaptive digital amplitude and delay pre-matching

Taixia Shi[a,b], and Yang Chen[a,b,*]

[a] Shanghai Key Laboratory of Multidimensional Information Processing, East China Normal University, Shanghai, 200241, China
[b] Engineering Center of SHMEC for Space Information and GNSS, East China Normal University, Shanghai, 200241, China
[*] ychen@ce.ecnu.edu.cn

**ABSTRACT**
A photonics-assisted analog wideband RF self-interference (SI) cancellation and frequency downconversion approach for in-band full-duplex (IBFD) multiple-input multiple-output (MIMO) systems with adaptive digital amplitude and delay pre-matching is proposed based on a dual-parallel Mach–Zehnder modulator (DP-MZM). In each MIMO receiving antenna, the received signal, including different SI signals from different transmitting antennas and the signal of interest, is applied to one arm of the upper dual-drive Mach–Zehnder modulator (DD-MZM) of the DP-MZM, the reference signal is applied to the other arm of the upper DD-MZM, and the local oscillator signal is applied to the lower DD-MZM. The SI signals are canceled in the optical domain in the upper DD-MZM and the frequency downconversion is achieved after photodetection. To cancel the SI signals, the reference signal is constructed in the digital domain, while the amplitude and delay of the constructed reference are adjusted digitally by upsampling with high accuracy. Experiments are performed when two different SI signals are employed. The genetic algorithm and least-squares algorithm are combined with segmented searching respectively for the SI signal reconstruction and amplitude and delay pre-matching. A cancellation depth of around 20 dB is achieved for the 1-Gbaud 16 quadrature-amplitude modulation orthogonal frequency-division multiplexing signal.

**Keywords:** Adaptive control, in-band full-duplex, microwave photonics, MIMO, self-interference cancellation.

## 1. Introduction

As the application of RF technology becomes more and more widespread, the demand for spectrum resources has also increased dramatically. However, spectrum resources are limited. In-band full-duplex (IBFD) systems transmit and receive signals at the same time and same frequency band, which can greatly improve the spectrum efficiency [1], [2]. When the IBFD operation is employed, a very strong self-interference (SI) is inevitably introduced, which cannot be suppressed by a band-pass or notch filter. Thus, new ways for SI cancellation (SIC) are needed.

Since the SI signal is commonly very strong in power due to the very short distance between the transmitting and receiving antennas, approaches from the antenna domain, analog domain, and digital domain are always coupled to provide a significant amount of cancellation depth [3]. The digital domain SIC methods are low-cost and flexible. However, to avoid the saturation of the low-noise amplifier and analog-to-digital converter (ADC), the antenna domain SIC and the RF analog SIC are indispensable. There are lots of electrical-based analog SIC methods to cancel the SI signal by constructing a reference signal using the known transmitted signal. However, due to the inherent constraints of cutting-edge electronic technology, the electrical-based approaches have a restricted working frequency and bandwidth. Utilizing microwave photonic signal processing to implement SIC can get beyond these restrictions [4], [5], taking the unique advantages of high frequency, large bandwidth, low transmission loss, and electromagnetic interference immunity offered by modern photonics.

In recent years, photonics-assisted SIC has been widely studied [6], [7]. From the perspective of microwave photonic system architecture, the reported SIC systems can be roughly divided into two categories: 1) using one optical path; 2) using two separated optical paths. When two separated optical paths are used, two intensity modulators [8]–[10]/equivalent intensity modulators [11]/phase modulators [12] combined with a photodetector (PD), two modulators/modulated lasers combined with a balanced photodetector [13]–[20], or a modulator with two orthogonal polarization states combined with a PD [21], are jointly used for the SIC. When one optical path is employed [22]–[31], a single modulator and a PD are used together to achieve the SIC. The system is more compact and may have better stability using a single optical path. Besides the SIC function, the photonics-assisted SIC methods combining frequency downconversion [21], [24], [26], [27] and radio over fiber transmission [20], [22], [24], [26]–[28] are also extensively studied.

For practical application, real-time adaptive control of the reference amplitude and delay is required to adapt to the variation of the wireless channel, especially when multipath SI signals are considered. Therefore, various control methods are investigated based on photonics-assisted SIC methods. In [10] and [15], the amplitude and delay/phase of the reference signal were adaptively adjusted using the Nelder-Mead simplex algorithm with more than 60 iterations with the assistance of a semiconductor optical amplifier. To decrease the number of the iteration, the modified Hooke–Jeeves algorithm [16] and regular triangle (RT) algorithm [17] were applied to optimize the reference delay and amplitude via the tunable optical delay line (TODL) and variable optical attenuator. To achieve the SIC adaptive control with the signal of interest (SOI) [18], [19], the bit error rate of the SOI was used as the optimization parameter by using the RT algorithm. In [21], for adaptive control with more dimensions, the particle swarm optimization (PSO) algorithm was applied to optimize the phase, amplitude, and time delay of the reference signal for the photonic-assisted SIC.

Multiple-input multiple-output (MIMO) systems use antenna arrays to greatly increase the channel capacity or resist multipath fading [32], which can also be operated in the IBFD condition. When different antennas in the MIMO system transmit different signals for increasing the capacity, the SI signals are much more complicated compared with conventional IBFD systems. Nevertheless, the reported SIC methods [8]–[31] are not designed for this kind of IBFD MIMO system and cannot be applied to it directly. In [20], photonics-assisted SIC methods were proposed for IBFD MIMO systems. However, the adaptive control of time delay and amplitude is not investigated and the MIMO antennas transmit the same signal for resisting multipath fading but not different signals for increasing the capacity. Therefore, the implementation of SIC methods for IBFD MIMO systems with different antennas transmitting different signals is particularly desirable. Furthermore, most of the reported photonics-based SIC methods directly manipulate the amplitude and delay of the reference signals in the analog domain, which can be limited in efficiency due to the restricted tuning speed, for example, the motorized TODL. Therefore, it is worth studying the delay and amplitude adaptive searching and adjusting method for the photonics-assisted SIC of IBFD MIMO systems without the limitation of the adjustment speed, such as the motorized TODL.

In this paper, a photonics-assisted analog wideband RF SIC and frequency downconversion approach for IBFD MIMO systems with adaptive digital amplitude and delay pre-matching is proposed based on a dual-parallel Mach–Zehnder modulator (DP-MZM). The SI signals are canceled in the optical domain in a dual-drive Mach–Zehnder modulator (DD-MZM) of the DP-MZM and the frequency downconversion is achieved after photodetection. To the best of our knowledge, this is the first photonics-assisted SIC method with adaptive delays and amplitudes control for IBFD MIMO systems whose transmitting antennas have different data streams. To cancel the SI signals, the reference signal is constructed in the digital domain, while the amplitude and delay of the constructed reference are adjusted in the digital domain by upsampling with high accuracy. In the experiment, two different data streams from two transmitting antennas of the MIMO system are used as the SI signals. When only the direct-path SI signals are considered, the genetic algorithm is used for amplitudes and delays searching, and a cancellation depth of more than 20 dB is achieved for the 1-Gbaud 16 quadrature-amplitude modulation (16-QAM) orthogonal frequency-division multiplexing (OFDM) signal. When both the direct-path and multipath SI signals are considered, the least-squares (LS) algorithm is used to estimate the SI signal for constructing the reference signal, the delay is found by segmented search, the amplitude of the reference can be calculated according to the SI signal power, and a cancellation depth of more than 19 dB is achieved for the 1-Gbaud 16-QAM OFDM signal.

**2. Principle**

## 2.1 Photonics-assisted SIC

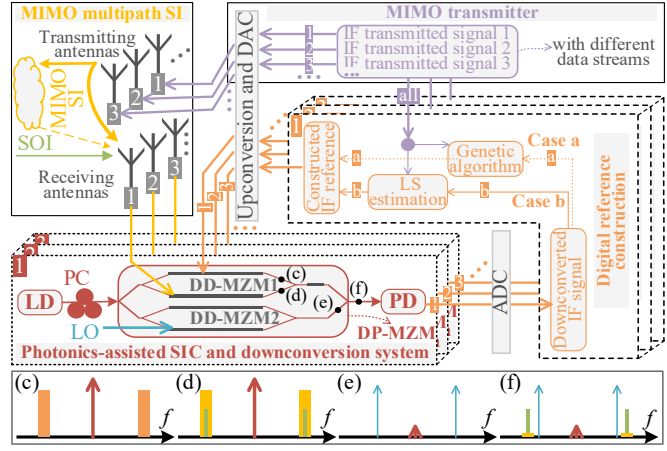

Fig. 1. Schematic of the proposed photonics-assisted SIC and frequency downconversion system. LD, laser diode; PC, polarization controller; DP-MZM, dual-parallel Mach–Zehnder modulator; DD-MZM, dual-drive Mach-Zehnder modulator; PD, photodetector; DAC, digital-to-analog converter; ADC, analog-to-digital converter; SI, self-interference; LO, local oscillator; SOI, signal of interest; LS, least-squares; Case a, only considering direct-path SI signals; Case b, considering both direct-path and multipath SI signals. (c)-(f) are the schematic diagrams of the signals at different locations in the system diagram.

The schematic of the proposed photonics-assisted RF SIC and frequency downconversion system for IBFD MIMO systems is shown in Fig. 1. The continuous-wave (CW) optical signal emitted by a laser diode (LD) is sent to a DP-MZM via a polarization controller (PC). The received signal, including both the SI signal and the SOI, is applied to the lower arm of DD-MZM1 in the DP-MZM. The reference signal for SIC is constructed in the digital domain with pre-matched amplitude and delay, converted to an analog reference signal, and then applied to the upper arm of DD-MZM1. A local oscillator (LO) signal is sent to one RF port of the DD-MZM2 for frequency downconversion. The two DD-MZMs are both biased at the minimum transmission point (MITP). Under these circumstances, the SI signals can be optically canceled in DD-MZM1, and the downconverted intermediate-frequency (IF) signal with SI suppressed can be obtained after photodetection. It should be emphasized that the SI scenario for an IBFD MIMO system is similar for each receiving antenna: the interference from every transmitting antenna is received. Thus, in this work, we only demonstrate the SIC after the signal is received by one of the receiving antennas. More specifically, we take two transmitting antennas as an example for subsequent analysis and experiments.

The SI signals (SI1 and SI2), the constructed reference signals (REF1 and REF2), the SOI, and the LO signal are expressed as

$$V_{s1}(t) = A_{s1} \cos[\omega_s(t-\tau_{s1}) + \phi_1], \tag{1}$$

$$V_{s2}(t) = A_{s2} \cos[\omega_s(t-\tau_{s2}) + \phi_2], \tag{2}$$

$$V_{r1}(t) = A_{r1} \cos[\omega_s(t - \tau_{r1}) + \phi_1], \tag{3}$$

$$V_{r2}(t) = A_{r2} \cos[\omega_s(t - \tau_{r2}) + \phi_2], \tag{4}$$

$$V_{SOI}(t) = A_3 \cos(\omega_s t + \phi_3), \tag{5}$$

$$V_L(t) = A_4 \cos(\omega_L t), \tag{6}$$

where $\omega_s$ is the center angular frequency of the SI signals, reference signals, and SOI, $\omega_L$ is the angular frequency of the LO signal, $A_{s1}$, $A_{s2}$, $A_{r1}$, $A_{r2}$, $A_3$, $A_4$, $\phi_1$, $\phi_2$, $\phi_1$, $\phi_2$, $\phi_3$, $\tau_{s1}$, $\tau_{s2}$, $\tau_{r1}$, and $\tau_{r2}$ are the amplitudes, phases, and delays of the corresponding signals. To simplify the following analysis, it is assumed that

$$\theta_{s1} = \omega_s(t - \tau_{s1}) + \phi_1, \tag{7}$$

$$\theta_{s2} = \omega_s(t - \tau_{s2}) + \phi_2, \tag{8}$$

$$\theta_{r1} = \omega_s(t - \tau_{r1}) + \phi_1, \tag{9}$$

$$\theta_{r2} = \omega_s(t - \tau_{r2}) + \phi_2, \tag{10}$$

$$\theta_3 = \omega_s t + \phi_3, \tag{11}$$

$$\theta_4 = \omega_L t. \tag{12}$$

Thus, these signals can be expressed as $A_{s1}\cos(\theta_{s1})$, $A_{s2}\cos(\theta_{s2})$, $A_{r1}\cos(\theta_{r1})$, $A_{r2}\cos(\theta_{r2})$, $A_3\cos(\theta_3)$, and $A_4\cos(\theta_4)$, respectively. It is assumed that the optical signal from the LD is $\exp(j\omega_c t)$. When both DD-MZM1 and DD-MZM2 are biased at MITP, the optical signals from the lower and upper arms of DD-MZM1 and DD-MZM2 can be written as

$$\begin{aligned} E_{1-l}(t) &= \frac{1}{2}\exp[j\omega_c t + jm_{s1}\cos(\theta_{s1}) + jm_{s2}\cos(\theta_{s2}) + jm_3\cos(\theta_3)] \\ &\approx \frac{1}{2}\exp(j\omega_c t)[J_0(m_{s1})J_0(m_{s2})J_0(m_3) \\ &\quad + 2jJ_1(m_{s1})J_0(m_{s2})J_0(m_3)\cos(\theta_{s1}) \\ &\quad + 2jJ_0(m_{s1})J_1(m_{s2})J_0(m_3)\cos(\theta_{s2}) \\ &\quad + 2jJ_0(m_{s1})J_0(m_{s2})J_1(m_3)\cos(\theta_3)], \end{aligned} \tag{13}$$

$$\begin{aligned} E_{1-u}(t) &= \frac{1}{2}\exp[j\omega_c t + jm_{r1}\cos(\theta_{r1}) + jm_{r2}\cos(\theta_{r2}) + \pi] \\ &\approx -\frac{1}{2}\exp(j\omega_c t)[J_0(m_{r1})J_0(m_{r2}) \\ &\quad + 2jJ_1(m_{r1})J_0(m_{r2})\cos(\theta_{r1}) \\ &\quad + 2jJ_0(m_{r1})J_1(m_{r2})\cos(\theta_{r2})], \end{aligned} \tag{14}$$

$$E_{2-l}(t) = \frac{1}{2}\exp[j\omega_c t + jm_4\cos(\theta_4)] \qquad (15)$$
$$\approx \frac{1}{2}\exp(j\omega_c t)[J_0(m_4) + 2jJ_1(m_4)\cos(\theta_4)],$$

$$E_{2-u}(t) = \frac{1}{2}\exp(j\omega_c t + \pi), \qquad (16)$$

where $J_n(\cdot)$ is the $n$th-order Bessel function of the first kind, $m_i = \pi A_i/V_\pi$ ($i=s1$, $s2$, $r1$, $r2$, 3, 4) are the modulation indices, and $V_\pi$ is the half-wave voltage of the DP-MZM. When small-signal modulation ($m_i \ll 1$, $J_0(m_i) \approx 1$, $J_1(m_i) \ll J_0(m_i)$) is employed, only the first-order optical sidebands are taken into account in the derivation, and the optical signal from the DP-MZM can be written as

$$\begin{aligned}E_{DP-MZM}(t) &\approx \frac{j}{2}\exp(j\omega_c t)[J_1(m_{s1})\cos(\theta_{s1}) \\ &+ J_1(m_{s2})\cos(\theta_{s2}) - J_1(m_{r1})\cos(\theta_{r1}) \\ &- J_1(m_{r2})\cos(\theta_{r2}) + J_1(m_3)\cos(\theta_3) \\ &+ J_1(m_4)\cos(\theta_4)].\end{aligned} \qquad (17)$$

In this article, the reference is pre-matched in the digital domain, so $\tau_{s1} = \tau_{r1}$, $\tau_{s2} = \tau_{r2}$, $A_{s1} = A_{r1}$, and $A_{s2} = A_{r2}$ established and the SI signals can be canceled in the optical domain. The SI-free optical signal from the DP-MZM can be derived as

$$E_{DP-MZM}(t) = \frac{j}{2}\exp(j\omega_c t)[J_1(m_3)\cos(\theta_3) + J_1(m_4)\cos(\theta_4)]. \qquad (18)$$

Then, the optical signal from the DP-MZM is detected by the PD with a responsivity of $R$, and the downconverted IF signal from the PD can be expressed as

$$i_{PD-IF}(t) = \frac{R}{4}J_1(m_3)J_1(m_4)\cos[(\omega_s - \omega_L)t + \phi_3]. \qquad (19)$$

As can be seen, the SI signals are canceled and the SI-free SOI is downconverted to the IF band.

Note that, in the above theoretical analysis, SI1 and SI2 represent two different SI signals from two different MIMO antennas, so we only consider the two direct-path SI signals, and the multipath SI signals from the two antennas are not taken into consideration. Indeed, when more direct-path SI signals or multipath SI signals are also considered, the conclusion of the above analysis is still valid, but the derivation process is more complex.

To realize good cancellation performance, the optimal time delay and amplitude of the constructed reference should be found. The amplitude attenuation factor of the reference signal can be determined by the power ratio of the SI signal and the reference signal without attenuation [31], whereas a rough estimation of the proper time delay of the reference signal can be obtained by a cross-correlation process. In the cross-correlation process, both the received signal and the reference signal without delay adjustment are applied to DD-MZM1. After the cross-correlation of the transmitted signal and the signal from the ADC, two correlation peaks corresponding to the SI signal and the reference signal can be obtained from the curve of cross-correlation. Thus, a rough estimation of the proper time delay can be acquired by calculating the difference in the correlation peaks between the SI signal and the unadjusted reference signal [31]. However, the accuracy and efficiency of the search are limited inherently, and more accurate and efficient amplitude and time delay searching of the reference signal are needed. In this article, the segmented searching method is used for amplitude and time delay search, and the genetic algorithm is also introduced to estimate the amplitude and time delay of the reference signal.

2.2 Segmented delay and amplitude adjustment in the digital domain

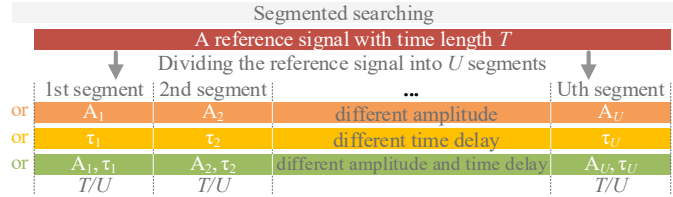

Fig. 2. Principle of the delay and amplitude segmented search in the digital domain.

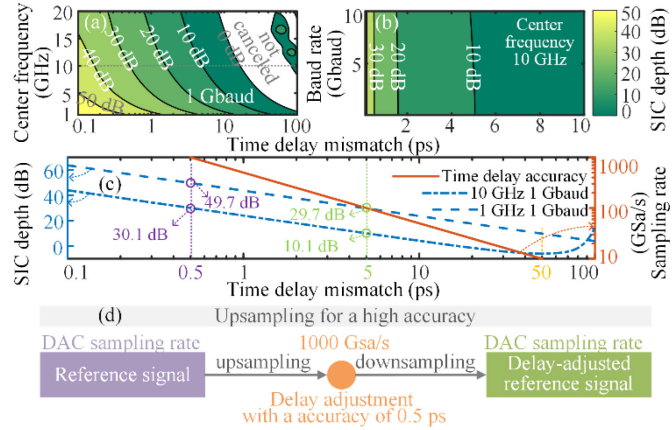

Fig. 3. Impact of the time delay mismatch on the SIC depth when the SI signal has different (a) center frequencies and (b) baud rates. (c) The SIC depth at different SI carrier frequencies and sampling rates. (d) Digital domain time delay adjustment accuracy improvement by upsampling.

One of the advantages of the delay and amplitude adjustment in the digital domain is the reference signal can be segmented for searching, which can greatly reduce the difficulty and improve the efficiency of the delay and amplitude searching. As shown in Fig. 2, a reference signal with a time length $T$ is divided into $U$ segments, and each segment can be set with a different amplitude and time delay. Therefore, the SIC performance of lots of different settings can be easily acquired for one searching iteration by measuring each corresponding segment residual power of the IF signal from the PD. Using segmented searching, it is very easy to find the optimal time delay and amplitude via a very small number of iterations. In addition, compared with the conventional methods which optimize the reference delay via the TODL, the segmented searching method in the digital domain require much fewer iterations and is not limited by the adjusting speed of the motorized device.

In our previous study [31], we experimentally verified the photonics-assisted SIC by adjusting the amplitude and delay in the digital domain. The SIC depth is limited by the time delay adjusting accuracy when the sampling rate used for delay adjusting is the same as the sampling rate of the digital-to-analog converter (DAC) especially when the carrier frequency of the SI signal is higher. To analyze the SIC performance influenced by the sampling rate of the signal without considering other mismatches, a simulation is performed with a digital sampling rate of 1000 Gsa/s. The SI signal is a 16-QAM OFDM signal. Fig. 3(a) shows that when the carrier frequency of the SI signal is higher than 5 GHz and the time delay mismatch is larger than 10 ps, the SIC depth is always less than 10 dB under a baud rate of 1 Gbaud. When the carrier frequency of the SI signal is further increased, a smaller time delay mismatch is required to achieve the same SIC depth. Fig. 3(b) shows that when the carrier frequency of the SI signal is 10 GHz, the SIC depth has no significant difference at different baud rates when only the time delay mismatch is considered. The time delay adjustment accuracy is determined by the signal sampling rate in digital signal processing (DSP). If the sampling rate is $f_s$, the time delay accuracy is $0.5/f_s$. For instance, when the digital sampling rate is 1000 Gsa/s, the time delay adjustment accuracy is 0.5 ps, and a cancellation depth of 30.1 dB can be achieved without other mismatches, as shown in Fig. 3(c). Fig. 3(d) shows the digital domain time delay adjustment accuracy improvement by upsampling. Thanks to the introduction of DSP, in the case of a limited DAC sampling rate, high-accurate delay matching can be achieved by upsampling. In practical applications, the rough time delay can be first found through cross-correlation, and then a more precise time delay can be searched using the segmented searching method in conjunction with upsampling.

2.3 Delays and amplitudes searching by genetic algorithm

In this work, different antennas in the MIMO system transmit different signals for increasing the capacity. First, a simpler case is considered: Only the direct-path SI signals from different transmitting antennas are taken into account. Therefore, the optimal values of $A_{r1}$, $A_{r2}$, $\tau_{r1}$, and $\tau_{r2}$ should be found. In this case, intelligent optimization algorithms, such as the genetic algorithm, are suitable to solve this kind of multidimensional optimization problem. Before using the genetic algorithm, the rough time delays and amplitudes were determined by the cross-correlation process and the power of the SI signal. The searching range of the genetic algorithm is set according to the obtained rough time delays and amplitudes. Thus, the multidimensional SIC problem is defined as follows [33]: Given a 4-dimensional variable vector **b** = {$A_{r1}$, $A_{r2}$, $\tau_{r1}$, $\tau_{r2}$} in the solution space **B** which corresponds to the search range, find a vector that maximizes the SIC depth. The designed procedure of the genetic algorithm for SIC optimal parameters searching is given as follows [33]:

Step 1: Set $g = 1$. Create the first population $G_1$ by randomly generating $N_p$ group solutions, and apply the first population to the segmented reference. Evaluate the SIC performance in $G_1$ according to each corresponding segment IF signal power from the PD.

Step 2: Crossover: Choose the solution which has the best SIC performance and the other half of the top solutions from $G_g$ based on the SIC performance and generate $N_p$ offspring by using a crossover operator, and add the generated offspring to $Q_g$.

Step 3: Mutation: Each solution in $Q_g$ is mutated with a predetermined mutation rate of $P_m$.

Step 4: Fitness assignment: Calculate the SIC depth of each solution in $Q_g$.

Step 5: Selection: select the top $N_p$ solutions from $Q_g$ and $G_g$ based on their SIC depths and copy them to $G_{g+1}$.

Step 6: If the predefined SIC depth is achieved, terminate the search and apply the best solutions to the SIC system, else, set $g = g+1$ and go to Step 2.

2.4 Constructing the reference signal using LS estimation

When both the direct path and multipath SI signals are considered for IBFD MIMO systems, the problem is more complicated and the intelligent optimization algorithms discussed above require much more iterations, resulting in an exponential increase in the algorithm complexity. Under these circumstances, the reference signal for SIC can be more efficiently constructed in the digital via the LS algorithm.

The vector representation of the received IF signal in one receiving antenna can be expressed as

$$\mathbf{y} = \mathbf{V}_s + \mathbf{V}_{SOI}, \tag{20}$$

where $\mathbf{V}_S$ is the IF SI signal, and $\mathbf{V}_{SOI}$ is the IF SOI signal. For the channel estimation of the SI signal, the received signals with $N$ samples are needed and can be written as

$$\mathbf{y} = \begin{bmatrix} y(n) & y(n+1) & \cdots & y(n+N-1) \end{bmatrix}. \tag{21}$$

The channel estimation $\hat{\mathbf{h}}$ by the LS algorithm can be estimated as [34]

$$\hat{\mathbf{h}} = \left(\mathbf{\Psi}^H \mathbf{\Psi}\right)^{-1} \mathbf{\Psi}^H \mathbf{y}, \tag{22}$$

where

$$\mathbf{\Psi} = \begin{bmatrix} \mathbf{\Psi}_1 & \mathbf{\Psi}_2 & \cdots & \mathbf{\Psi}_{NT} \end{bmatrix}, \tag{23}$$

$$\mathbf{\Psi}_j = \begin{bmatrix} x_j(n) & x_j(n-1) & \cdots & x_j(n-M) \\ x_j(n+1) & x_j(n) & \cdots & x_j(n-M+1) \\ \vdots & \vdots & \ddots & \vdots \\ x_j(n+N-1) & x_j(n+N-2) & \cdots & x_j(n-M+N-1) \end{bmatrix}, \tag{24}$$

where $j=1, 2,\ldots, NT$, $NT$ is the number of transmitting antennas, $x_j(n)$ corresponds to the IF signal of the transmitted signal of the $j$-th transmitting antenna, and $M$ is the order of the LS algorithm. Thus, the total reference for analog SIC can be constructed by the estimated channel information and the known transmitted signal. The constructed IF reference that includes all the SI information can be expressed as

$$\hat{\mathbf{r}} = \mathbf{\Psi}\hat{\mathbf{h}}. \tag{25}$$

After the reference construction by the LS estimation, the amplitude and time delay information of the constructed reference is required. The amplitude of the constructed reference via the LS algorithm can be determined by the SI signal power. The rough time delay of the total reference can be first found through cross-correlation, and then a more precise time delay can be searched using the segmented searching method in conjunction with the upsampling. It should be noted that the reference signal constructed by the LS algorithm contains all the SI information and has the relative amplitude and delay information of all the SI paths. Therefore, there is no need to find the delay and amplitude of each SI path and only a single delay and amplitude of the constructed reference needs to be found.

## 3. Experiment and results

### 3.1 Experimental setup

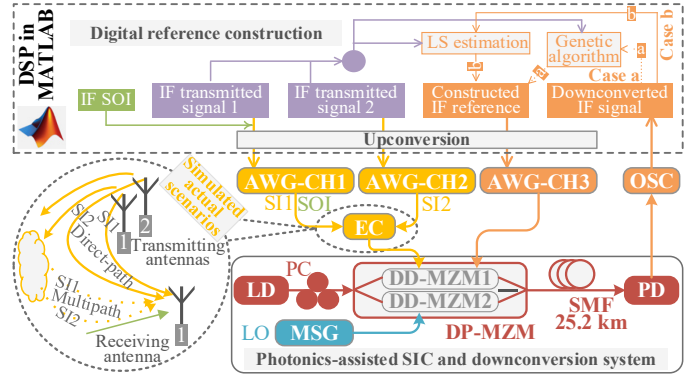

Fig. 4. Experimental setup of the proposed photonics-assisted SIC and frequency downconversion system. AWG, arbitrary waveform generator; CH, channel; EC, electrical coupler; MSG, microwave signal generator; OSC, oscilloscope; SMF, single-mode fiber.

An experiment is carried out to verify the proposed method based on the setup shown in Fig. 4. The CW optical signal from an LD (ID Photonics CoBriteDX1-1-C-H01-FA) with a wavelength of 1550.053 nm and a power of 15.5 dBm is sent to a DP-MZM. An arbitrary waveform generator (AWG, Keysight M8195A, 64 Gsa/s) is used to simulate and generate the signals in the real-world IBFD MIMO system. In the experiment, two SI signals with different data streams from two transmitting antennas and their multipath SI signals are considered to simulate the SI signals in an IBFD MIMO system with increased capacity, while in the receiving end, only the SIC following one receiving antenna is investigated because other receiving ends have the same SIC system structure and principle. The SI signals from the two transmitting antennas are generated from two different channels of the AWG, while the digitally constructed reference signal is generated from the third channel of the AWG, all with an output peak-to-peak amplitude of 1 V. One direct-path SI signal and its multipath SI signals are denoted as SI1, while the other direct-path SI signal and its multipath SI signals are denoted SI2. In addition, the SOI is generated along with the SI signals in one channel. The LO signal is generated by a microwave signal generator (MSG, Agilent 83630B) with a power of 20 dBm. The SI signals and SOI from the AWG are combined at an electrical coupler (EC, Narda 4456-2, 2~18 GHz, –3 dB) and applied to the lower arm of the DD-MZM1 in the DP-MZM (Fujitsu FTM7960EX301), whereas the digital pre-matched reference is applied to the upper arm of the DD-MZM1. The LO signal is sent to one RF port of the DD-MZM2. Both DD-MZM1 and DD-MZM2 are biased at MITP, and the SI signals are canceled in the optical domain. Then, the optical signal from the DP-MZM is injected into a PD (Nortel Networks PP-10G), and the IF electrical signal from the PD is captured by an oscilloscope (OSC, R&S RTO2032, 10 Gsa/s). The digitized IF signal from the OSC is processed in a computer using Matlab and then the constructed digital reference signal from Matlab is downloaded to the AWG to generate the analog reference signals for SIC.

3.2 Reference delay searching accuracy improvement by digital upsampling

Because the amplitude of the constructed reference can be adjusted according to the SI signal power and the accuracy is high enough to guarantee good cancellation performance, the segmented searching method proposed in Fig. 2 is only used to obtain the accurate reference delay in conjunction with cross-correlation in the experiment.

First, the time delay searching with higher accuracy by upsampling in the digital domain is investigated. In this study, the experiment is performed based on Fig. 4 but only employs the direct-path SI signal from one transmitting antenna, which is a 16-QAM OFDM signal with a center frequency of 9 GHz and a baud rate of 1 Gbaud, respectively. The LO frequency is set to 8 GHz. The amplitude attenuation factor is calculated to be around 0.51 according to the SI signal power. Fig. 5(a) shows that the cross-correlation delay difference between the transmitted signal and the combined signal of the unadjusted reference signal and SI signal is 4.9 ns. For more accurate time delay searching of the reference signal, the segmented searching method with upsampling to 1000 Gsa/s in the digital domain is employed, and Fig. 5(b) and (c) show the corresponding results. When the 4-μs reference signal is divided into 200 segments from 4800~4999 ps, the optimal delay is found to be 4880 ps which is actually not the optimal delay value. As shown in Fig. 5(c-ii), only a 13.7-dB cancellation depth is obtained in this case. It should be noted that the SIC depth has a small difference obtained from the power difference and the spectra, especially for a short time length signal owning to the noise influences. Thus, the searching range is enlarged to 4700~5099 ps, and the 4-μs reference signal is divided into 400 segments. Fig. 5(c-i) shows that the optimal delay value is 4768 ps and the cancellation depth is greatly improved to 27.3 dB. Fig. 5(d) shows the comparison of the SIC performance with and without upsampling in the delay searching. It can be seen that the best SIC depth is around 27.3 dB when upsampling (1000 Gsa/s) is employed, whereas the best SIC depth is around 14 dB when the delay is adjusted under the sampling rate of the AWG (64 Gsa/s). When the center frequency of the SI signals is changed to 10 GHz and the LO frequency is changed to 9 GHz, a result similar to that in Fig. 5(d) is obtained, as shown in Fig. 5(e).

There are some interferences in the spectra of Fig. 5(d) and (e). To more clearly distinguish these interferences from the residual SI signals after SIC, the output of the AWG is enabled without downloading any signal. It can be clearly observed that the frequencies of the interferences caused by the AWG are mainly at 2, 1, 2, and 3 GHz, respectively, when the frequency of the LO signal is changed from 8 to 11 GHz with a step of 1 GHz, as shown in Fig. 6. Thus, these interferences should not be considered when calculating the SIC depth from the spectra.

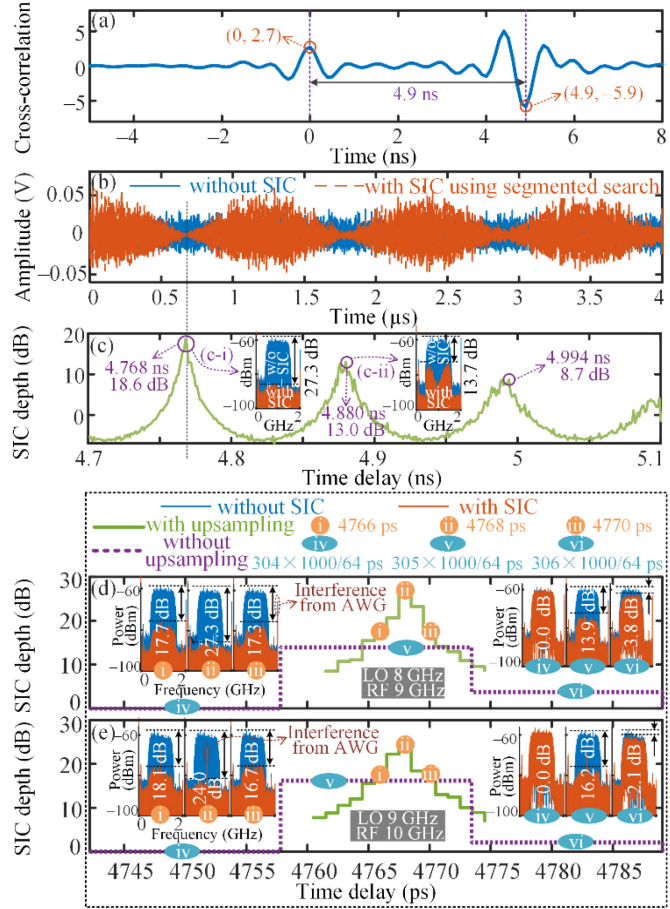

Fig. 5. Demonstration of the delay searching accuracy improvement by segmented searching and digital upsampling. (a) Cross-correlation curve. (b) Waveforms without SIC and with SIC using segmented searching. (c) SIC depth of segmented searching. SIC performance with and without upsampling when the frequency of LO signal is (d) 8 GHz and (e) 9 GHz.

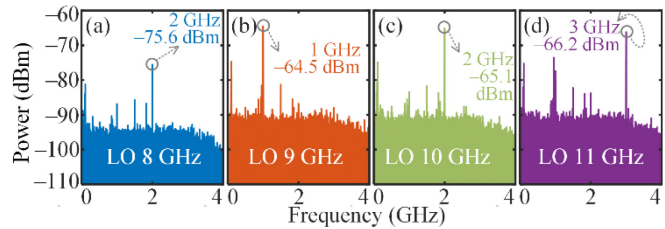

Fig. 6. The electrical spectra of the interference from the PD introduced by the AWG.

### 3.3 SIC for IBFD MIMO systems considering only direct-path SI signals

For IBFD MIMO systems, if only the direct-path SI signals are taken into account, in general, it can be considered that the SI signals from the direct path are fixed and without variation when the beams of the transmitting antenna are not scanned and the transmitting power is not changed. Under these circumstances, the amplitudes and delays of the direct-path SI signals are also fixed and unchanged, so these parameters can be obtained in advance by sequentially transmitting signals on each transmitting antenna and measuring the amplitude and delay from each transmitting antenna to each receiving antenna.

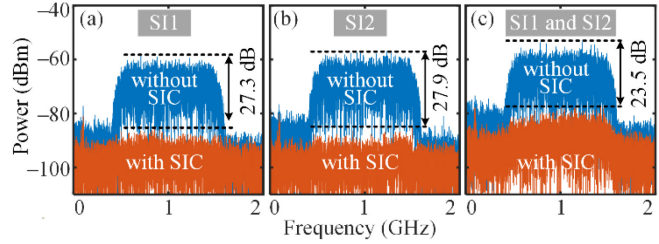

Fig. 7. IBFD MIMO SIC when (a) only SI1, (b) only SI2, and (c) both SI1 and SI2 are applied to the DD-MZM1.

Here, in our experiment, two direct-path SI signals from two channels of the AWG are sequentially sent to the SIC system. One SI signal is the same as used in Section 3.2, and the other SI signal is with a new set of delay and amplitude. These two SI signals have completely different data streams. The time delays and amplitudes of the two direct-path SI signals are searched using the methods discussed in Section 3.2, which are 4768 ps and 0.51, and 3828 ps and 0.53, respectively. Fig. 7(a) and (b) show the spectra of the two SI signals with and without SIC. For both SI signals, a cancellation depth of around 27.5 dB can be achieved. After getting the amplitudes and delays of the direct-path SI signals in advance, the SIC system can still work when all the transmitting antennas transmit signals simultaneously, as long as the transmitting power is fixed and the beams of the transmitting antennas are not scanned. Fig. 7(c) shows the spectra of the SI signals with and without SIC when the two direct-path SI signals are enabled simultaneously. A cancellation depth of about 23.5 dB is obtained.

The method discussed above is applicable, but with restrictions: The transmitting beam and power cannot be changed, which means the delays and amplitudes measured in advance will not be applicable in future advanced IBFD MIMO systems, that is, it is necessary to track the delay and amplitude changes of the direct-paths in real-time adaptively.

3.4 Adaptive tracking of the SI signal delays and amplitudes variation based on genetic algorithm

To solve the problem mentioned above, a genetic algorithm as discussed in Section 2.3 is used to adaptively track the SI signal delay and amplitude variation. Specifically, two direct-path SI signals with different data streams are considered in this study as a 4-dimensional optimization problem. The center frequency and baud rate of the SI signal are first set to 9 GHz and 1 Gbaud, respectively, the LO frequency is set to 8 GHz, and the SOI is still not applied.

In order to reduce the search complexity of the genetic algorithm, an extra stage (Stage 1) is used to estimate the search ranges of time delays and amplitude attenuation factors of two reference signals, which can also be implemented by using the cross-correlation method and SI signal power. As shown in Fig. 8, when the received signal consisting of two different direct-path SI signals (SI1 and SI2) are cross-correlated with two reference signals (REF1 and REF2), respectively, the rough SI signal delays are around 4.9 and 4.0 ns. Therefore, the ranges of the time delays of the two reference signals are set to 4700~5100 ps and 3800~4200 ps, respectively. The amplitudes of the two reference signals can be searched from 0 to 1. However, because the delay difference between two SI signals from two transmitting antennas is not very large, the power difference between the two SI signals is also not very large. Thus, although the amplitude attenuation factor obtained using the total SI power is not accurate due to the mixing of the SI signals, the value can still be used as a reference value to search the two amplitudes of the SI signals in a smaller range rather than from 0 to 1. In this experiment, the range of the amplitude attenuation factors of REF1 and REF2 are set to 0.24~0.74 with a resolution of 0.01.

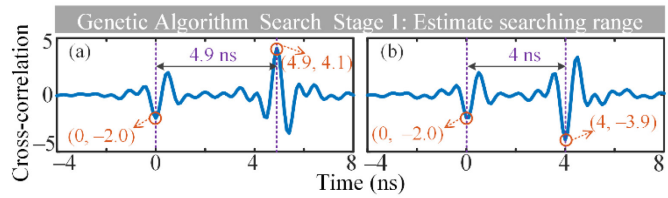

Fig. 8. Cross-correlation curves between the transmitted signal and the combined signal of the unadjusted reference signal and (a) SI1 and (b) SI2.

After determining the search ranges, genetic algorithm search Stage 2 for large range search is further employed. In this stage, both the 4-μs REF1 and 4-μs REF2 are divided into 160 segments. 152 segments of them are used for genetic algorithm search and the other 8 segments are used for experimental synchronization convenience, so the population number $N_P$ is 152. The mutation rate $P_m$ is set to 0.1. The search range determined in Stage 1 is used in Stage 2, and eleven iterations are performed to search for the optimal time delays and amplitude attenuation factors. Fig. 9(a) shows the distribution of the time delays and amplitude attenuation factors of 152 individuals in the population of the first and eleventh iterations. It can be seen that these parameters are uniformly distributed throughout the search area owing to the random generation in the first iteration. However, in the eleventh iteration, most of the parameters converge to some specific value due to heredity, and some parameters are not converged owing to the crossover and mutation operation. Fig. 9(b) shows that the optimal time delays and amplitude attenuation factors changed with eleven iterations in five searches. It can be seen that the converge ranges of the time delay and the amplitude attenuation factor of REF1 and REF2 are 4766~4771 ps, 3827~3831 ps, 0.42~0.56, and 0.46~0.57, respectively. Fig. 9(c) shows the SIC depth is increased with the iterations in both five searches. Fig. 9(d) shows the enveloped electrical spectra of the downconverted IF signal from the PD captured by the OSC when using the optimal delay and amplitude setting of some specific iterations in five searches. It can be seen that at least 16.11 dB

SIC depth can be achieved after eleven iterations.

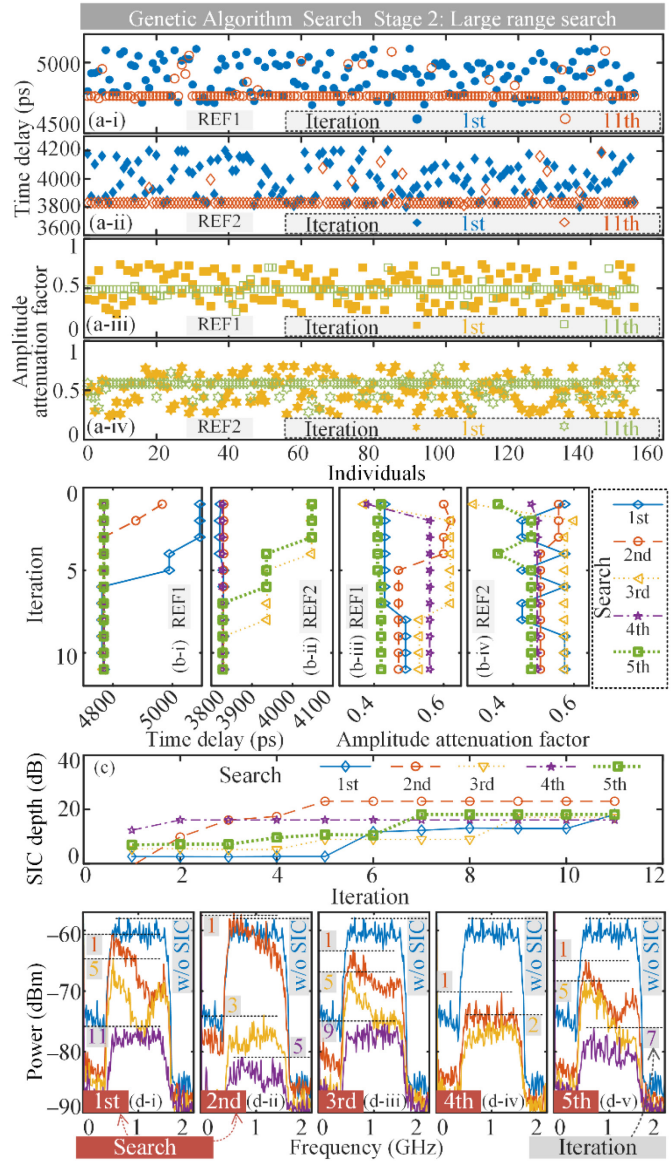

Fig. 9. Genetic algorithm search Stage 2. (a) Distribution of the time delays and amplitude attenuation factors of individuals. (b) Optimal delay and amplitude setting of eleven iterations in five searches. (c) SIC depth trends with the iterations in five searches. (d) Some enveloped electrical spectra of the downconverted IF signal from the PD in five searches.

For a more accurate search, the genetic algorithm search is executed once again in Stage 3. In this stage, the search is reduced based on the converged values in Stage 2. Fig. 10(a) shows that the optimal time delays and amplitude attenuation factors changed with eleven iterations in five searches in Stage 3. It can be seen that the converge ranges of the time delays of REF1 and REF2 are 4767~4769 ps and 3828~3829 ps. The amplitude attenuation factors converge to 0.46~0.51 and 0.44~0.49. It can be seen that the range of the convergence distribution is reduced by half compared to the genetic algorithm search Stage 2. Fig. 10(b) shows that the SIC depth is increased with the iterations in both five searches. Fig. 10(c) shows the enveloped electrical spectra of the downconverted IF signal from the PD when using the optimal delay and amplitude setting of some specific iterations in five searches. It can be seen that a SIC depth of more than 12.34 dB can be obtained in the first iteration. After eleven iterations, a minimum SIC depth of 20.74 dB is obtained in the third search, whereas a maximum SIC depth of 26.63 dB is obtained in the fourth search. Compared with the result in Stage 2, the SIC depth can be increased by at least 4.6 dB and at most 10.5 dB.

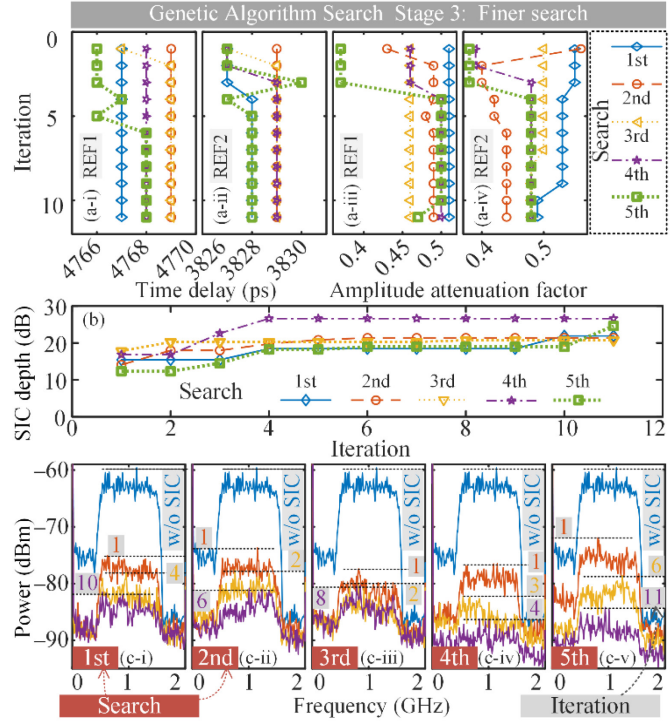

Fig. 10. Genetic algorithm search Stage 3. (a) Optimal delay and amplitude setting of eleven iterations in five searches. (b) SIC depth trends with the iterations in five searches. (c) Some enveloped electrical spectra of the downconverted IF signal from the PD in five searches.

To further study the performance of the genetic algorithm search in other signal baud rate, the baud rate of the SI signals is changed to 500 Mbaud. Using the genetic algorithm to search the optimal time delays and amplitude attenuation factors, the population number $N_P$ and the mutation rate $P_m$ are not changed. The search range is determined by using the same methods for the case of SI signals with 1 Gbaud. Fig. 11(a) shows the SIC depth is increased with the iterations in both three searches in Stage 2. The SIC depths of the three searches are 18.7, 15.1, and 14.3 dB, respectively, as shown in Fig. 11(b). Fig. 11(c) shows the SIC depth is increased with the iterations in both three searches in Stage 3, and the converge ranges of the time delay and the amplitude attenuation factor of REF1 and REF2 are 4766~4767 ps, 3827~3831 ps, 0.56~0.59, and 0.57~0.59, respectively. The SIC depths of the three searches are 25.1, 21.4, and 23.5 dB, respectively, as shown in Fig. 11(d). It can be seen that the SIC depth can also be increased by 7.1 dB from Stage 2 to Stage 3 when the baud rate of the SI signals is 500 Mbaud.

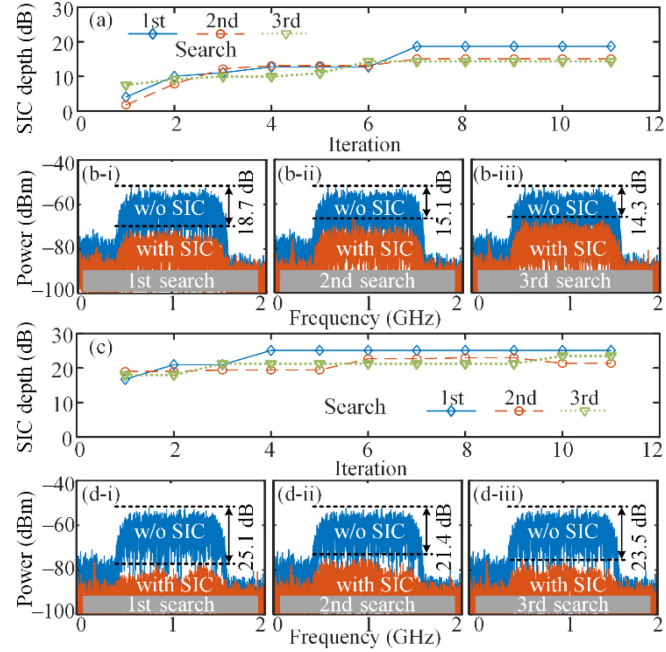

Fig. 11. Genetic algorithm-based searches when the baud rate is 500 Mbaud. SIC depth trends with the iterations in three searches in (a) Stage 2 and (c) Stage 3. The electrical spectra of the downconverted signal from the PD in three searches in (b) Stage 2 and (d) Stage 3.

3.5 SIC for IBFD MIMO systems by LS estimate for both direct-path and multipath SI signals

When both the direct-path and multipath SI signals are considered in IBFD MIMO systems, the problem is more complicated. To avoid too many iterations and too much algorithm complexity by using intelligent optimization algorithms, the reference signal for SIC can be constructed digitally via the LS algorithm as discussed in Section 2.4.

To achieve SIC via the LS estimation, the optimal time delay and amplitude attenuation factor also need to be found for the reference signal after the SI signals, including the direct-path SI signals from different antennas and their multipath SI signals, are estimated and the corresponding reference signal is constructed. The amplitude attenuation factor of the reference signal can also be determined according to the SI signal power, and the optimal time delay can be found by the cross-correlation and segmented searching combining upsampling, as discussed in Section 3.2. It should be pointed out here that the reference signal constructed by the LS algorithm contains all information of all the SI paths and only a single delay and amplitude of the constructed reference needs to be found. The difference between this study and that in Section 3.2 is the reference: The reference in this study is constructed by taking all the direct-path SI signals and multipath SI signals into consideration, while that in Section 3.2 is constructed by only considering one direct-path SI signal.

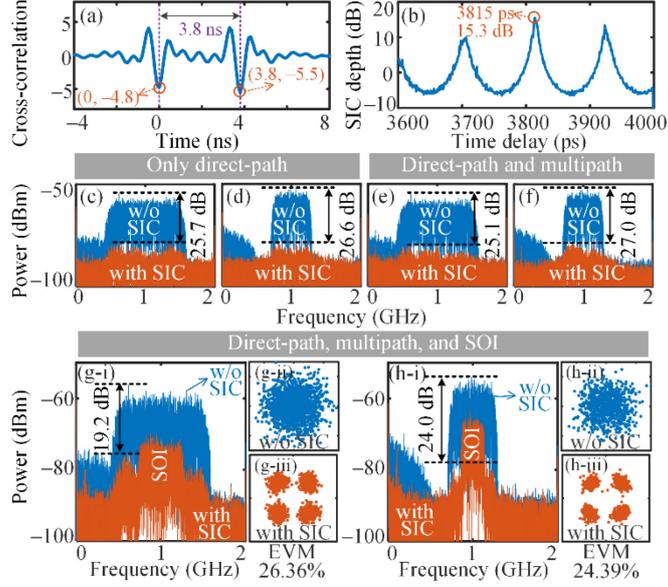

Fig. 12. SIC for IBFD MIMO systems via LS estimation. (a) Cross-correlation curve. (b) SIC depth of segmented searching. Electrical spectra of the downconverted IF signal from the PD with and without SIC when only the direct-path SI signals with (c) 1 and (d) 0.5 Gbaud are employed, and when direct-path and multipath SI signals with (e) 1 and (f) 0.5 Gbaud are employed. Electrical spectra and constellation diagrams when the SOI, the direct-path SI signals, and multipath SI signals with (g) 1 and (h) 0.5 Gbaud are employed.

An example of a time delay search for the constructed reference signal is shown in Fig. 12(a) and (b). Fig. 12(a) shows that the correlation delay difference between the SI signals and the unadjusted reference signal is 3.8 ns when the center frequencies of the SI signals and LO signal is 9 and 8 GHz and the baud rate of the 16-QAM OFDM SI signals is 1 Gbaud. For more accurate time delay searching of the constructed reference signal, the segmented searching method with upsampling to 1000 Gsa/s is used. The 4-μs constructed reference signal is divided into 400 segments, and the optimal delay can be found to be 3815 ps, as shown in Fig. 12(b).

TABLE I
MULTIPATH DELAY AND GAIN SETTINGS

|  | SI1 multipath relative to SI1 direct-path | SI2 multipath relative to SI2 direct-path |
|---|---|---|
| Delay | 8/13/15 ns | 7/15/17 ns |
| Gain | –10/–12/–15 dB | –10/–12/–15 dB |

Then, various cases are further studied when using LS constructed reference for the analog SIC of IBFD MIMO systems. The center frequencies of the LO signal and the SI signals are 8 and 9 GHz and the baud rate of the SI signals is 0.5 or 1 Gbaud. When the received signal includes only direct-path SI signals, the SIC depths are 25.7 dB and 26.6 dB for 1- and 0.5-Gbaud SI signals, as shown in Fig. 12(c) and (d). When the received signal includes both direct-path and multipath SI signals, the setting of the multipath SI signals is shown in Table I and kept unchanged in the following study. The

SIC depths are 25.1 dB and 27.0 dB for 1- and 0.5-Gbaud SI signals which are similar to the case of having only direct-path SI signals, as shown in Fig. 12(e) and (f). When the SOI is taken into consideration, the SOI is set to be a quadrature phase-shift keying (QPSK) signal, the center frequency of the SOI is the same as that of the SI signals, and the baud rate of the SOI is set to half that of the SI signals for convenience of observation. These settings are also not changed in the subsequent investigation. Fig. 12(g) and (h) show that the SOI is submerged in the spectrum by the SI signals and the constellation diagrams of the SOI are chaotic when the SIC is not employed. When the SIC is enabled, the SOI is no longer submerged by the SI signals, and the SIC depths of 1- and 0.5-Gbaud SI signals are 19.2 dB and 24.0 dB, respectively. The constellation diagrams of the SOI can be clearly distinguished and the corresponding error vector magnitudes (EVMs) of the SOI are 26.36% and 24.39%. It can be seen that, when the SOI is added, the SIC performance decreases a little but can still effectively suppress the complex SI signals.

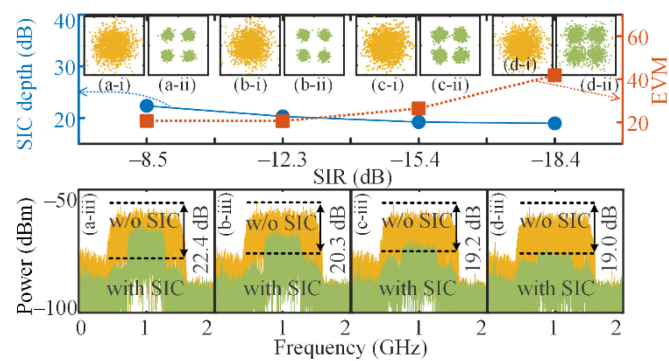

Fig. 13. IBFD MIMO analog SIC by LS estimation when the SIR is (a) –8.5 (b) –12.3 (c) –15.4 and (d) –18.4 dB.

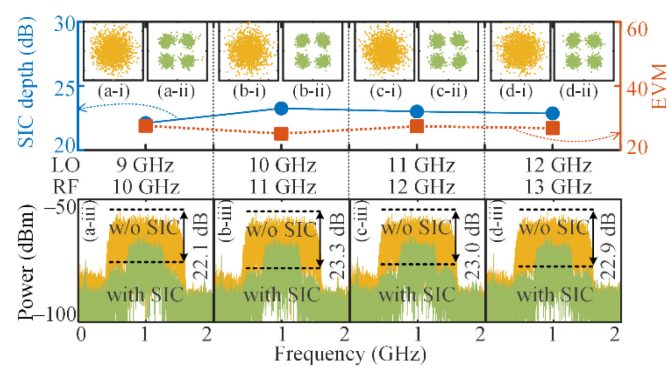

Fig. 14. IBFD MIMO analog SIC by LS estimation at different LO and RF frequencies.

Then, the SOI is still included in the following experiments. Fig. 13 shows the performance of the IBFD MIMO analog SIC by LS estimation in different signal-to-interference ratio (SIR) conditions. With the decrease of the SIR, SIC depths of around 20 dB are achieved. Nevertheless, the EVMs of the SOI become worse and the constellation diagrams of the SOI tend to be chaotic with the decrease of the SIR. This phenomenon is mainly because the power of SOI decreases with the decrease of the SIR, and the influence of noise and residual SI signals is more obvious when the power of SOI decreases.

Afterward, the frequency tunability of the IBFD MIMO analog SIC system using LS estimation is also demonstrated with the results shown in Fig. 14. As can be seen, the SIC depth, the EVM of the SOI, and the constellation diagrams of the SOI after SIC are very similar at different LO and RF frequencies.

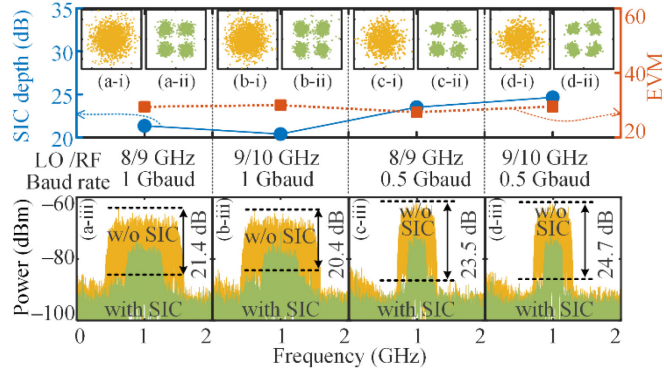

Fig. 15. IBFD MIMO analog SIC by LS estimation at different LO and RF frequencies and baud rates when antenna remoting is employed by inserting a section of 25.2-km SMF.

When antenna remoting is used in the MIMO system, it is highly desirable that the signal received in the antenna can be transmitted back to the central office for further signal processing [24]. Therefore, the performance of the IBFD MIMO analog SIC system is further studied by inserting a section of 25.2 km single mode fiber (SMF) for antenna remoting applications. The center frequency of the SI signals and SOI is set to 10 or 11 GHz, the center frequency of the LO signal is set to 9 or 10 GHz, and the baud rate of the SI signals is set to 1 or 0.5 Gbaud. Experimental results are shown in Fig. 15. The SIC performance is very similar to that without inserting the 25.2-km SMF, and SIC depths are around 21 dB and 24 dB for the 1- and 0.5-Gbaud SI signals. The EVMs and constellation diagrams of the SOI after SIC are also similar to that without fiber transmission. It should be noted that since the power attenuation of the used 25.2-km SMF is measured to be around 4.6 dB, there is also a power attenuation of the downconverted IF signal which is measured to be around 8.8 dB. Furthermore, thanks to the benefits of the IF fiber transmission and optical domain SIC [24] used in the proposed system, the SIC is not influenced by the fiber dispersion and the reception of the SOI is also not influenced by the fiber dispersion-induced fading effect.

**4. Discussion**

4.1 Cross-correlation and segmented searching

To ensure the accuracy of the delay searching process, i.e., the cross-correlation and segmented searching, some settings should be considered. For a rough estimation of the time delay by the cross-correlation process, the baud rate of the signals determines the width of the correlation peaks (the time resolution of the estimation determined by the reciprocal of the baud rate) while the IF carrier frequency determines the internal variation of the envelope of the correlation peaks. Fig. 16 shows the cross-correlation results when the center frequencies and baud rates of the IF signals are 1, 1, 1, 1, 0.5, 0.25 GHz and 1, 0.5, 0.25, 0.1, 0.5, 0.25 Gbaud, respectively. The set time delay is around 4.768 ns. It can be seen that when the baud rate of the IF signal is higher, it is much easier to get a more accurate delay value. If the reciprocal of the baud rate is smaller than the delay value, as shown in Fig. 16(d), the delay estimation via cross-correlation will have much worse accuracy. Nevertheless, the IF frequency is also very important for the delay estimation, especially when the set time delay is very close to the reciprocal of the baud rate because when the time delay approaches the width of the correlation peak, the low IF frequency will cause the two correlation peaks to overlap with each other as shown in Fig. 16(f).

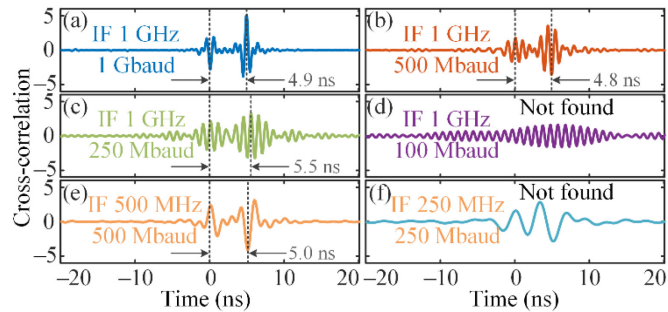

Fig. 16. Cross-correlations at different baud rates and IF center frequencies.

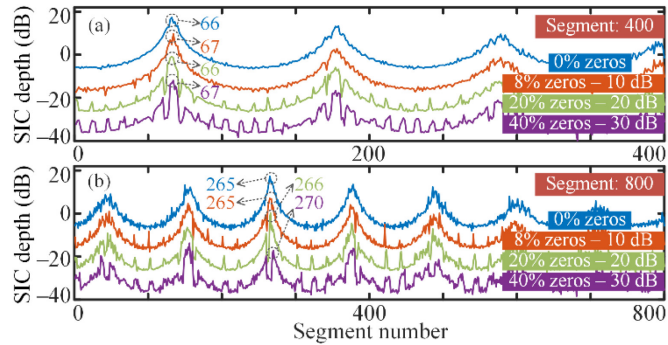

Fig. 17. SIC depth of segmented searching under different lengths of zero padding when the 4-μs reference signal is divided into (a) 400 or (b) 800 segments.

In the segmented searching process, the searching accuracy is influenced by the length of zero padding of the OFDM SI signals. An experiment is further conducted by using SI signals with different lengths of zero padding, the results are shown in Fig. 17. In the segmented searching, the reference signal is divided into 400 and 800 segments. It can be seen that with the increase of the percentage of zero padding in the OFDM SI signals, the searching curve turns noisier and the searching accuracy will be degraded. When the percentage of zero padding is high, if the optimal value found by the segmented searching method cannot get a good SIC performance, we can do a small-scale search near this value to approach a better result.

4.2 Comparison of SIC by genetic algorithm and LS algorithm

The adaptive control of the amplitudes and delays of the reference signals is a multidimensional optimization problem in the IBFD MIMO systems whose transmitting antennas have different data streams at the same carrier frequency. To solve this problem, the genetic algorithm and LS algorithm are used for the SIC of the IBFD MIMO systems, and both two methods achieve similar SIC performance.

Using the genetic algorithm combining the segmented searching method can directly solve the multidimensional optimization problem. In our experiment, the 4-dimensional optimization is implemented by the genetic algorithm. The experiment is carried out by combining the online Matlab-based signal processing, waveform generation from the AWG, and the analog optoelectronic system. Only 4-dimensional optimization is realized because the efficiency of the experiment is highly limited by the data transmission and processing speed in Matlab and the waveform download time. The efficiency and number of dimensions can be increased by implementing the genetic algorithm and signal generation on hardware [35].

By using the LS estimation to first estimate the SI channel response, this multidimensional problem is reduced to a 2-dimensional problem. Thus, compared to the SIC method using the genetic algorithm, the complexity of that using the LS algorithm is much reduced. However, in order to track the SI channel variations, the SI channel information needs to re-estimate the channel every once in a while. Therefore, there is an interruption to analog SIC when capturing the SI signals for the LS estimation. Thus, compared to the SIC method using the LS algorithm, the genetic algorithm is easier to track the best parameters for SIC without interruption to the analog SIC when the searching of the genetic algorithm is converged to a small searching range.

## 5. Conclusion

In summary, we have demonstrated a photonics-assisted analog wideband RF SIC and frequency downconversion approach for IBFD MIMO systems with adaptive digital amplitude and delay pre-matching. To the best of our knowledge, this is the first photonics-assisted SIC method with adaptive delays and amplitudes control for IBFD MIMO systems whose transmitting antennas have different data streams. The time delay and amplitude were searched and adjusted in the digital domain by segmented searching and upsampling with improved accuracy. When only the direct-path SI signals were considered, the genetic algorithm was used for amplitudes and delays searching, and a cancellation depth of more than 20 dB was achieved for the 1-Gbaud 16-QAM OFDM signal. When both the direct path and multipath SI signals were considered, the LS algorithm was used to estimate the SI signal for constructing the reference signal, the delay was found by segmented search, and the amplitude of the reference was determined according to the SI signal power, and a cancellation depth of more than 19 dB was achieved for the 1-Gbaud 16-QAM OFDM signal. The approach proposed in this paper is a pioneering exploration for SIC in IBFD MIMO systems and is expected to be a solution for complex SI problems in future IBFD MIMO systems.

**Acknowledgements**

This work was supported by the National Natural Science Foundation of China [grant number 61971193]; the Natural Science Foundation of Shanghai [grant number 20ZR1416100] ; the Science and Technology Commission of Shanghai Municipality [grant number 18DZ2270800].

**References**

[1] Z. Zhang, X. Chai, K. Long, A. V. Vasilakos, and L. Hanzo, "Full duplex techniques for 5G networks: Self-interference cancellation, protocol design, and relay selection," *IEEE Commun. Mag.*, vol. 53, no. 5, pp. 128–137, May 2015.

[2] A. Sabharwal, P. Schniter, D. Guo, D. W. Bliss, S. Rangarajan, and R. Wichman, "In-band full-duplex wireless: challenges and opportunities," *IEEE J. Sel. Areas Commun.*, vol. 32, no. 9, pp. 1637–1652, Sep. 2014.

[3] K. E. Kolodziej, B. T. Perry, and J. S. Herd, "In-band full-duplex technology: techniques and systems survey," *IEEE Trans. Microw. Theory Techn.*, vol. 67, no. 7, pp. 3025–3041, Jul. 2019.

[4] J. Yao, "Microwave photonics," *J. Lightw. Technol.*, vol. 27, no. 3, pp. 314–335, Feb. 2009.

[5] J. Capmany, J. Mora, I. Gasulla, J. Sancho, J. Lloret, and S. Sales, "Microwave photonic signal processing," *J. Lightw. Technol.*, vol. 31, no. 4, pp. 571–586, Feb. 2013.

[6] X. Han, X. Su, S. Fu, Y. Gu, Z. Wu, X. Li, and M. Zhao, "RF self-interference cancellation by using photonic technology," *Chin. Opt. Lett.*, vol. 19, no. 7, Jun. 2021, Art. No. 073901.

[7] V. J. Urick, M. E. Godinez, and D. C. Mikeska, "Photonic assisted radio-frequency interference mitigation," *J. Lightw. Technol.*, vol. 38, no. 6, pp. 1268–1274, Mar. 2020.

[8] J. Suarez, K. Kravtsov, and P. R. Prucnal, "Incoherent method of optical interference cancellation for radio-frequency communications," *IEEE J. Quantum Electron.*, vol. 45, no. 4, pp. 402–408, Apr. 2009.

[9] J. Chang, and P. R. Prucnal, "A novel analog photonic method for broadband multipath interference cancellation," *IEEE Microw. Wireless Compon. Lett.*, vol. 23, no. 7, pp. 377–379, Jul. 2013.

[10] M. P. Chang, C. Lee, B. Wu, and P. R. Prucnal, "Adaptive optical self-interference cancellation using a semiconductor optical amplifier," *IEEE Photon. Technol. Lett.*, vol. 27, no. 9, pp. 1018–1021, May 2015.

[11] W. Zhou, P. Xiang, Z. Niu, M. Wang, and S. Pan, "Wideband optical multipath interference cancellation based on a dispersive element," *IEEE Photon. Technol. Lett.*, vol. 28, no. 8, pp. 849–851, Apr. 2016.

[12] X. Han, B. Huo, Y. Shao, C. Wang, and M. Zhao, "RF self-interference cancellation using phase modulation and optical sideband filtering," *IEEE Photon. Technol. Lett.*, vol. 29, no. 11, pp. 917–920, Jun. 2017.

[13] M. P. Chang, M. Fok, A. Hofmaier, and P. R. Prucnal, "Optical analog self-interference cancellation using electro-absorption modulators," *IEEE Microw. Wireless Compon. Lett.*, vol. 23, no. 2, pp. 99–101, Feb. 2013.

[14] Y. Zhang, L. Li, S. Xiao, M. Bi, L. Huang, L. Zheng, and W. Hu "EML-based multi-path self-interference cancellation with adaptive frequency-domain pre-equalization," *IEEE Photon. Technol. Lett.*, vol. 30, no. 12, pp. 1103–1106, Jun. 2018.

[15] M. P. Chang, E. C. Blow, J. J. Sun, M. Z. Lu, and P. R. Prucnal, "Integrated microwave photonic circuit for self-interference cancellation," *IEEE Trans. Microw. Theory Techn.*, vol. 65, no. 11, pp. 4493–4501, Nov. 2017.

[16] L. Huang, Y. Zhang, S. Xiao, L. Zheng, and W. Hu, "Real-time adaptive optical self-interference cancellation system for in-band full-duplex transmission," *Opt. Commun.*, vol. 437, no. 15, pp. 259-263, Apr. 2019.

[17] L. Zheng, Y. Zhang, S. Xiao, L. Huang, J. Fang, and W. Hu, "Adaptive optical self-interference cancellation for in-band full-duplex systems using regular triangle algorithm," *Opt. Exp.*, vol. 27, no. 4, pp. 4116–4125, Feb. 2019.

[18] L. Zheng, S. Xiao, Z. Liu, M. P. Fok, J. Fang, H. Yang, M. Lu, Z. Zhang, and W. Hu, "Adaptive over-the-air RF self-interference cancellation using a signal-of-interest driven regular triangle algorithm," *Opt. Lett.*, vol. 45, no. 5, pp. 1264–1267, 2020.

[19] Z. Zhang, L. Zheng, S. Xiao, Z. Liu, J. Fang, and W. Hu, "Real-time IBFD transmission system based on adaptive optical self-interference cancellation using the hybrid criteria regular triangle algorithm," *Opt. Lett.*, vol. 46, no. 5, pp. 1069–1072, Mar. 2021.

[20] X. Yu, J. Ye, L. Yan, T. Zhou, X. Zou, and W. Pan, "Photonic-assisted multipath self-interference cancellation for wideband MIMO radio-over-fiber transmission," *J. Lightw. Technol.*, vol. 40, no. 2, pp. 462–469, Jan. 2022.

[21] X. Hu, D. Zhu, L. Li, and S. Pan, "Photonics-based adaptive RF self-interference cancellation and frequency downconversion," *J. Lightw. Technol.*, vol. 40, no. 7, pp. 1989–1999, Apr. 2022.

[22] Y. Zhang, S. Xiao, H. Feng, L. Zhang, Z. Zhou, and W. Hu, "Self-interference cancellation using dual-drive Mach-Zehnder modulator for in-band full-duplex radio-over-fiber system," *Opt. Exp.*, vol. 23, no. 26, pp. 33205–33213, Dec. 2015.

[23] X. Han, B. Huo, Y. Shao, and M. Zhao, "Optical RF self-interference cancellation by using an integrated dual-parallel MZM," *IEEE Photon. J.*, vol. 9, no. 2, Apr. 2017, Art. No. 5501308.

[24] Y. Chen, and S. Pan, "Simultaneous wideband radio-frequency self-interference cancellation and frequency downconversion for in-band full-duplex radio-over-fiber systems," *Opt. Lett.*, vol. 43, no. 13, pp. 3124–3127, Jul. 2018.

[25] X. Li, Y. Zhang, L. Huang, L. Deng, M. Cheng, S. Fu, M. Tang, and D. Liu, "Optimized self-interference cancellation based on optical dual-parallel MZM for co-frequency and co-time full duplex wireless communication under nonlinear distortion and emulated multipath effect," *Opt. Exp.*, vol. 27, no. 26, pp. 37286–37297, Dec. 2019.

[26] S. Zhu, M. Li, N. Zhu, and W. Li, "Photonic radio frequency self-interference cancellation and harmonic down-conversion for in-band full-duplex radio-over-fiber system," *IEEE Photon. J.,* vol. 11, no. 5, Oct. 2019, Art. No. 5503110.

[27] Y. Chen, and J. Yao, "Photonic-assisted RF self-interference cancellation with improved spectrum efficiency and fiber transmission capability," *J. Lightw. Technol.*, vol. 38, no. 4, pp. 761–768, Feb. 2020.

[28] Y. Chen, "A photonic-based wideband RF self-interference cancellation approach with fiber dispersion immunity," *J. Lightw. Technol.*, vol. 38, no. 17, pp. 4618–4624, Sep. 2020.

[29] L. Zheng, Z. Liu, S. Xiao, M. P. Fok, Z. Zhang, and W. Hu, "Hybrid wideband multipath self-interference cancellation with an LMS pre-adaptive filter for in-band full-duplex OFDM signal transmission," *Opt. Lett.*, vol. 45, no. 23, pp. 6382–6385, Dec. 2020.

[30] M. Han, T. Shi, and Y. Chen, "Digital-assisted photonic analog wideband multipath self-interference cancellation," *IEEE Photon. Technol. Lett.*, vol. 34, no. 5, pp. 299–302, Mar. 2022.

[31] T. Shi, M. Han, and Y. Chen, "Photonics-assisted wideband RF self-interference cancellation with digital domain amplitude and delay pre-matching," *Optik*, Vol. 250, no. 1 Jan. 2022, Art. No. 168343.

[32] L. Lu, G. Y. Li, A. L. Swindlehurst, A. Ashikhmin, and R. Zhang, "An overview of massive MIMO: Benefits and challenges," *IEEE J. Sel. Topics Signal Process.*, vol. 8, no. 5, pp. 742–758, Oct. 2014.

[33] A. Konaka, D. W. Coitb, and A. E. Smith, "Multi-objective optimization using genetic algorithms: A tutorial," *Reliab. Eng. Syst. Safety*, vol. 91, no. 9, pp. 992–1007, Sep. 2006.

[34] D. Korpi, L. Anttila, and M. Valkama, "Nonlinear self-interference cancellation in MIMO full-duplex transceivers under crosstalk" *EURASIP J. Wireless Commun. Netw.*, vol. 2017, no. 1, Feb. 2017, Art. no. 24.

[35] P. Fernando, S. Katkoori, D. Keymeulen, R. Zebulum, and A. Stoica, "Customizable FPGA ip core implementation of a general-purpose genetic algorithm engine," *IEEE Trans. Evol. Comput.*, vol. 14, no. 1, pp. 133–149, Feb. 2010.